\documentclass[twocolumn,showpacs,prl]{revtex4}
\usepackage{amssymb}
\usepackage{graphicx}
\usepackage{dcolumn}
\usepackage{bm}

\begin{document}

\title{Electron Self-injection in Multidimensional Relativistic Plasma Wakefields }
\author{I.~Kostyukov}
\email{kost@appl.sci-nnov.ru}
\author{E.~Nerush}
\affiliation{Institute of Applied Physics, Russian Academy of Science, 46 Uljanov St. 603950 Nizhny Novgorod, Russia}
\author{A.~Pukhov, V.~Seredov}
\affiliation{Institut fur Theoretische Physik I, Heinrich-Heine-Universitat Duesseldorf,
40225 Duesseldorf, Germany}
\date{\today}

\begin{abstract}
We present an analytical model for electron self-injection in nonlinear,
multidimensional plasma wave excited by short laser pulse in the
bubble regime or by short electron beam in the blowout regime. In this
regimes, which are typical for electron acceleration in the last experiments, 
the laser radiation pressure or the electron
beam charge pushes out background plasma electrons 
forming a plasma cavity - bubble - with a huge ion charge. The plasma
electrons can be trapped in the bubble and accelerated by the
plasma wakefields up to very high energies. The model predicts the condition for
electron trapping and the trapping cross section in terms of the bubble
radius and the bubble velocity. The obtained results are in a good agreement with results of
3D PIC simulations. 
\end{abstract}

\maketitle

Plasma-based charged particle acceleration is now a fast
developing area of science. It has attracted much attention due
to the recent experimental breakthrough in generation of
quasimonoenergetic, dense and short bunches of 
relativistic electrons with up to GeV energies in the
laser-driven acceleration \cite{Mangles2004,Geddes2004,Faure2004,Leemans2006} 
and due to the energy doubling of a $42$ GeV
electron beam \cite{Blumenfeld2007}. The high-gradient acceleration is
provided by very strong electromagnetic fields in a plasma wake excited by short
laser pulse or electron beam. The plasma electrons are
radially expelled by ponderomotive force of the laser pulse or by
Lorentz force from the space charge of the electron beam that leads to
formation of the plasma cavity - bubble with unshielded ions
inside. This is the bubble \cite{Pukhov2002} or blowout 
\cite{Rosenzweig1991} regime of laser-plasma and beam-plasma
interaction. It is important that a small part of the background
plasma electrons can be trapped in the bubble providing the electron
self-injection. The bubble velocity is close to the speed
of light and the trapped relativistic electrons can be continuously
accelerated in the plasma fields up to a very high energy.  

The electron self-injection is the key phenomenon of the last 
laser-plasma acceleration experiments \cite{Mangles2004,Geddes2004,Faure2004,Leemans2006}. 
It is a crucial factor for the quality of the accelerated electron beam. Many
applications ranging from x-ray free electron lasers to
electron-positron colliders require high quality electron beams with a 
very low emittance and energy spread. The one-dimensional model based
on the Hamiltonian formalism has been developed to study electron
trapping in relativistic plasma wave
\cite{Esarey1995,Schroeder2006}. Several optical and plasma techniques
have been proposed to enhance the electron self-injection in plasma
wakefield like the collision of two counter-propagating laser pulses
\cite{Esarey1997,Faure2006}, density transitions
\cite{Bulanov1998,Suk2001,Esirkepov2006} etc. In the ultrahigh
intensity regime the radial force from the driver and 
from the plasma fields strongly affect electron motion and
the electron dynamics become complex and multidimensional. 
 The trajectory of the trapped electrons
starts at the bubble front, bends around the cavity and becomes caught
at the trailing edge of the bubble. Evidently, such trajectories
cannot be described in the framework of a one-dimensional model. 

Despite the great interest in plasma-based electron acceleration  
there is little theory of self-injection. In Ref.~\cite{Lu20062,Lu2006,Tzoufras2008} 
a multidimensional model of a relativistic plasma wake including 
beam loading is proposed, however the electron self-injection was not 
considered. The electron self-injection has been mainly studied numerically
\cite{Zhidkov2004,Kostyukov2004,Lu2007}. It was found in Ref.~\cite{Lu2007} 
that a sufficiently large bubble, $R > 4$, can trap  plasma 
electrons. Here $R$ is the bubble radius normalized to $c / \omega_p$, 
$\omega_p = \sqrt{4\pi n_0 e^2 /m}$ is the (non-relativistic) plasma 
frequency for the background electron density $n_0$, $e$ and $m$, 
are the charge and mass of the electron, respectively. 
A different condition for electron self-injection has been proposed
in Ref.~\cite{Kostyukov2004}: $R > \gamma _0$, where $\gamma _0$ is the bubble gamma-factor.
Unfortunately, principal limitations of a numerical approach make the accuracy 
and the validity range of the obtained conditions unclear and do not
provide an insight in the self-injection physics. 
However, the numerical approach has its natural limitations 
in accuracy and validity conditions of the obtained results and 
does not provide insight in the self-injection physics.

In this paper, we 
present an analytical model for self-injection and 
check it by direct measurements of model parameters in 3D particle-in-cell (PIC)  simulations.
For simplicity we assume that the bubble has a spherical shape with
 large radius $R \gg 1$ that is typical for relativistic laser
pulses. The numerical simulations \cite{Kostyukov2004} and  
theoretical analysis \cite{Lu2006} demonstrate that the bubble shape
is close to the spherical one for relativistically intense laser pulse and 
when the number of the trapped electrons is not too large. It can be shown 
\cite{Kostyukov2004,Lu2006} that the space-time distribution of the
electromagnetic field inside the spherical bubble expressed in cylindrical
coordinates is $E_{x}= (1+V) \xi /4 $, $E_\bot = 
- B_{\theta}=  \mathbf{ r}_\bot /4$, where $\xi =  x-Vt$, and $V \simeq 1$ is the bubble
velocity. We use dimensionless units, normalizing the time to $\omega
_{p}^{-1}$, the lengths to $c/\omega _{p}$, the velocity to $c$, the
electromagnetic fields to $mc\omega _{p}/|e|$, and the electron
density, $n$, to $n_{0}$. The distribution 
of the electromagnetic field inside the plasma cavity can be also expressed 
through vector and scalar potential \cite{Kostyukov2004} $A_{x}=r^2 /8$, 
$\mathbf{A}_{\perp }=0$ with gauge $A_{x}=-\varphi $, where $r^{2}= 
\xi ^{2}+y^{2}+z^2$ is the distance to the bubble center.  

We assume that the electron trajectory is plane ($z=0$) because of the axial
symmetry of the wakefield and the driver. The driver field is neglected
because it is weak at the region where trapping occurs (far behind
the laser pulse or electron beam). The field of the trapped particles
is also neglected since we assume that the number of the trapped particle is not
too large. The electron dynamics in the bubble is governed by the
Hamiltonian \cite{Kostyukov2004} $ H = \sqrt{1+\left[\Pi _x + 
A_x (\xi,y) \right]^2 + \Pi _y^2} - V \Pi _x - \varphi \left(\xi ,y \right)$,
where $\mathbf {\Pi }$ is the canonical momentum of the electron 
and $\xi $, $y$ are coordinates canonically conjugated to $\mathbf {\Pi }$.  
As the Hamiltonian is not the function of time then it is integral of motion 
$H = \text{const} $.  The Hamiltonian equations are: 
\begin{eqnarray}
\frac{dp_{x}}{dt} & = & -(1+V) \frac{\xi }{4} + \frac{p_y}{\gamma } \frac{y}{4} ,
\; 
\frac{dp_{y}}{dt} =  - \left( 1 + \frac{p_x}{\gamma } \right) \frac{y}{4}, 
\label{Hameq2}
\\
\frac{d\xi }{dt} &=&\frac{p_x}{\gamma }-V,\; 
\frac{dy}{dt}=\frac{p_y}{\gamma },  
\label{Hameq3}
\end{eqnarray}

\noindent where $\gamma =\left( 1+p_{x}^{2}+p_{y}^{2} \right)^{1/2}$ is the relativistic
$\gamma-$ factor of the electron, $\mathbf{p} = \mathbf{v}\gamma = 
\mathbf{\Pi  + A} (\xi ,y ) $ is the electron momentum. 

\begin{figure}[tbp]
\includegraphics[width=8cm]{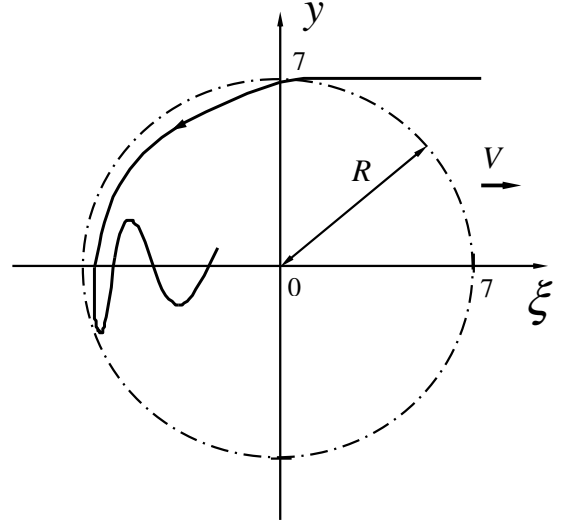}  
\caption{ Trajectory of the trapped (solid line) and untrapped electron 
(dashed line) calculated by numerical
solution of equation of electron motion and the bubble border
(dashed circle). The coordinates are given in $c/\omega _{p}$. }
\label{fig1}
\end{figure}

We assume that the electron is located initially at the bubble border $\mathbf{p} = 0 $,
$y = R$ and $\xi  = 0$ at $t=0$. It follows from the initial condition that 
\begin{equation}
H = \gamma -V p_x- (1+V) \varphi \left( \xi ,y\right) = 1 + (1+V)\frac{R^2}{8} ,
\label{H}
\end{equation}

\noindent where $\varphi (r = \infty) = \varphi (r = R) = - R^2 / 8  $. 
We solve the electron motion equations numerically. To
be more realistic and in accordance with PIC simulations we include an electron sheath around the plasma
cavity, which screens the bubble ion field in the surrounding
plasma. We model the electromagnetic fields inside the bubble as 
follows $E_x = f(r) \xi / 2 $, $E_y = - H_z = f(r) y / 4 $, where  
$ f(r) =  [\tanh (R/d-r/d) - 1]/2$, $d $ is the width of the electron sheath. 
The similar distribution of the wake field with $d<1$ is observed in the 
3D PIC simulation \cite{Kostyukov2004,Lu2007}. 
The typical trajectory of a trapped electron is
shown in Fig.~\ref{fig1} for $R=7$, $d=0.3$ and $\gamma _0 = 4$. 
The sheath does not affect the 
electron trajectory in the bubble when $d < 2$.  We
can conclude from the numerical solutions that the electron very soon becomes
ultrarelativistic so that $p_x \gg p_y \gg 1$.  The
electron undergoes betatron oscillations about the $\xi$-axis and slowly
moves along this axis ($|d\xi /dt|=|p_{x}/\gamma -V | \ll 1$). 
The condition $d \xi / d t \geq 0$ can be considered as a
necessary (but not sufficient) one for electron trapping. The electron leaves 
the bubble and is lost when $r > R$. The most critical instants of time 
when the electron can leave the bubble is $t = t_m > 0$ when 
$dy/dt = p_{y}=0$ and the electron excursion from the $\xi-$axis 
reaches its maximum. At this moment we can
write $\gamma -Vp_{x}\simeq p_{x}/\left( 2\gamma _{0}^{2}\right)
+1/\left( 2p_{x}\right) $, where $p_{x}\gg 1$ is assumed, $\gamma
_{0} = (1-V^2)^{-1/2} \gg 1$ is the gamma-factor of the bubble. The
square of the distance of the electron from the bubble center ($\xi
=0$, $y=0$) at $t=t_{m}$ is 
\begin{equation}
r_{m}^{2} \simeq 4+R^{2}-\frac{2p_{x}(t_{m})}{\gamma _{0}^{2}}-\frac{2}{p_{x}(t_{m})},  
\label{rmlin}
\end{equation}

\noindent where we use Eq.~(\ref{H}).

We change the variables $p_{y}=R^{2}P_{y}$, $p_{x}=R^{2}P_{x}$, $\xi =XR$, $y=YR$ 
and $t=Rs$. As a result Eqs.~(\ref{Hameq2}) and (\ref{Hameq3}) takes a form
\begin{eqnarray}
\frac{dP_{x}}{ds} & = & - \frac{X }{2} + \frac{P_y}{\sqrt{P_x^2 + P_y^2} } 
\frac{Y}{4} + O \left(R^{-4}, \gamma _0^{-2} \right),
\label{H1}
\\
\frac{dP_{y}}{ds} &= & - \left( 1 + \frac{P_x}{\sqrt{P_x^2 + P_y^2}} \right) 
\frac{Y}{4} + O \left(R^{-4} \right) , 
\label{H2}
\\
\frac{dX }{ds} &=&\frac{P_x}{\sqrt{P_x^2 + P_y^2} }-1 + O \left(R^{-4}, 
\gamma _0^{-2} \right) , 
\label{H3}
\\
\frac{dY}{ds} &=& \frac{P_y}{\sqrt{P_x^2 + P_y^2} } + O \left(R^{-4} \right) ,  
\label{H4}
\end{eqnarray}

\noindent where $P_{x}=P_{y}=0$ and $dP_{y}/ds=-1/4$, $dP_{x}/ds=0$ at $s=0$. 
In the zeroth order in $\gamma _{0}^{-2}$ and $R^{-4}$ Eqs.~(\ref{H1})-(\ref{H4}) 
do not depend on any parameters and can be numerically solved to find $s_{m}$ 
and $ t_{m}=Rs_{m}+O\left( \gamma _{0}^{-2} , R^{-2}\right) $: $ s_1\simeq 3.2$, 
$s_{2}\simeq 10.7$, $s_{3}\simeq 23.0$, .... 
Using Eq.~(\ref{rmlin}) the condition that electron leaves the bubble ($r_{m}>R$) 
can be written as follows 
\begin{equation}
2 >  \frac{R^2 P_x (s_m)}{\gamma _0^2} +O\left( \gamma _0^{-4} , R^{-2} \right)  .
\label{condlin}
\end{equation}

\noindent The electron leaves the bubble if there exists at least one value 
of $s_{m}$ when the condition (\ref{condlin}) is satisfied. We can take 
only $P_{x}\left( s_{1}\right) \simeq 1.1 $ since $P_{x}\left( s_{1}\right) < P_{x}
\left( s_{2}\right) <P_{x}\left( s_{3}\right) <...$. As a result we come 
to the condition for electron capture in the bubble 
\begin{equation}
\frac{\gamma _0}{R} \lesssim \frac{1}{ \sqrt{2}},
\label{cond}  
\end{equation}
 which is close to the condition obtained numerically in Ref.~\cite{Kostyukov2004}.

Now we estimate the effect of the impact parameter on the trapping condition.
If $\rho < R$ then electron first moves through the decelerating bubble field 
where $\xi > 0$. As a result the electron gains the negative momentum, $\Delta $, 
at $\xi = 0$ in contrast to the electron with $\rho = R $, 
which starts motion from position $\xi = 0 $ with $ \mathbf{P} =0 $.  
The deceleration leads to reduction of the longitudinal momentum at $s=s_1$ 
for electron with $\rho < R$ as compared to the electron with $\rho = R$. 
We estimate this reduction, $P_x (s_1, \rho = R) - P_x (s_1, \rho < R) $, as $\Delta $, 
that is close to the value calculated by the numerical integration of 
Eqs.~(\ref{Hameq2}), (\ref{Hameq3}). To calculate $\Delta $ we can use 
Eq.~(\ref{H}) and assume $1/R \ll R - \rho \ll R$.
As $p_y \simeq 0$ at $\xi = 0$, $y \simeq \rho$ then 
$H \simeq (1+R^4 \Delta ^2 )^{1/2} + R^2 \Delta + \rho ^2 / 4  =
1 + R^2 /4$, so that $\Delta = \nu (\nu +2/R^2)/ (2 \nu + 2/ R^2) \simeq \nu /2 $, 
where $\nu = (1 - \rho^2 / R^2)/4$. Therefore, 
$P_x (s_1, \rho \leq R) \simeq  P_x (s_1, \rho = R) - \nu / 2 $ and 
the condition for electron trapping (\ref{cond}) can be rewritten 
as follows  
\begin{eqnarray}
\rho_0 \lesssim \rho \lesssim R, 
\label{cond-gen1}
\\
\frac{ \rho _0^2}{R^2}  \simeq 1 -  8 \left( 1 - 2 \frac{\gamma _0^2}{R^2} \right).
\label{cond-gen2}
\end{eqnarray}%
\noindent 
Therefore the trapping cross-section $\sigma$ near the trapping threshold $
 \gamma _0  \simeq 2^{-1/2} R $ takes the form 
\begin{equation}
\frac{\sigma }{\pi R^{2}} = \int_{\rho _{0}}^{R}\frac{2\rho }{R^{2}}d\rho 
= 1-\frac{ \rho _0^2}{R^2}  \simeq 8 \left( 1 - 2 \frac{\gamma _0^2}{R^2} \right).  
\label{trap2}
\end{equation}%

\noindent
It follows from Eq.~(\ref{trap2}) that the trapping cross-section decreases as 
$\gamma _0 /R$ increases and the trapping stops at the threshold   
 $ \gamma _0  \simeq 2^{-1/2} R $.

\begin{figure}[tbp]
\includegraphics[width=6cm]{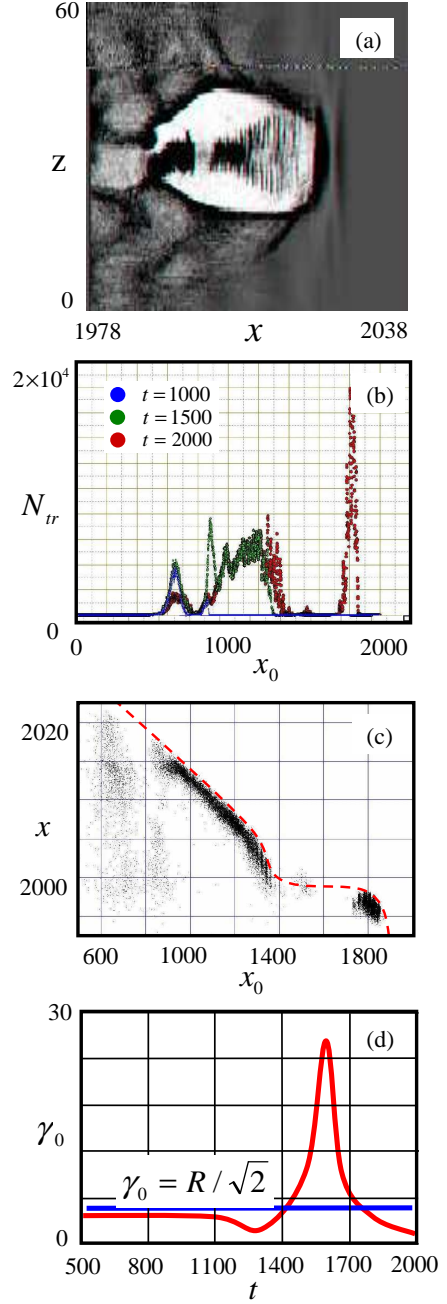}  
\caption{ 
(a) The distribution of the electron density in the plane $x-z$ is
calculated by 3D PIC simulation. (b) The number of the trapped
electron per laser period as function of $x_0$. (c) The current coordinate, $x$, of the
trapped electrons as function of $x_0$ and function approximating $x(x_0)$ (red dashed line). (d) The
gamma-factor of the bubble as function of $t$ estimated as $\gamma _0
\simeq \left( - 2 dx/dx_0 \right)^{-1/2}$ (red line) and the 
trapping condition $ \gamma _0 =  2^{-1/2} R$ (blue line).  The coordinates and time
are given in the laser wavelength and laser periods, the density is
 given in the critical plasma density, respectively.
 }
\label{fig2}
\end{figure}

The developed model abstracts from some important 
effects observed in numerical simulations: the bubble shape deformation
during laser pulse propagation and  
the bubble field enhancement at the bubble back due to electron sheet 
crossing \cite{Bulanov1997,Lu2007}. It follows from 3D PIC simulations that 
the bubble back typically moves slowly than the bubble front \cite{Kostyukov2004,Lu2007}. 
The bubble back gamma-factor should be used in Eq.~(\ref{cond}) as $\gamma _0$ 
since the electrons are trapped at the bubble back. 
Another effect is that the bubble field can be stronger at the bubble back 
than it follows from linear approximation for the bubble field 
in our model. To estimate the effect of field enhancement 
on the trapping we introduce the field enhancement factor $k$ so that 
$A_x=-\varphi=k r^2 /8$. It follows from the scalability analysis of 
Eqs.~(\ref{H1})-(\ref{H4}) that the coefficient at RHS of 
Eq.~(\ref{cond}) becomes $\sqrt{k/2}$. 
The considered effects can significantly affect electron self-injection. 
For parameters of simulations performed in Ref.~\cite{Lu2007}  
the minimal radius of the bubble with self-injection is reduced by 
a factor of $4$ when the bubble field enhancement 
($\sim 4$, see Fig.~3 in Ref.~\cite{Lu2007}) and the bubble back gamma-factor 
($\sim 9$, see Fig.~1 in Ref.~\cite{Lu2007}) 
are taken into account in Eq.~(\ref{cond}). The estimated radius 
is about $1.5$ times more 
than observed in the simulations. Further studies are needed to 
include the above mentioned effects in our model accurately.

We carry out a numerical simulation of laser-plasma interaction by
PIC VLPL3D code. The laser pulse is circularly
polarized and has the envelope $a=a_{0}\exp \left( -r^{2}/r_{L}^{2} \right)
\cos \left( \pi t/T_{L}\right) $, where $T_{L}=30fs$ is the pulse
duration, $r_{L}=9\mu m$ is the focused spot size and $a_{0}\equiv
eA/mc^{2}=1.5$ is the normalized vector potential, which corresponds
to the laser intensity $I=3 \times 10^{18}W/cm^{2}$, $\lambda = 0.82
\mu m$ is the laser wavelength. The plasma density is $n_{0}=1.16
\times 10^{19}cm^{-3}$ so that the parameters are close to that used in experiments
\cite{Faure2004}. The typical distribution of the electron density in
the bubble regime is shown in Fig.~\ref{fig2}(a) when laser pulse has
passed the distance $2000 \lambda$ in plasma. 
Because of the large number of trapped
electrons the field enhancement effect is suppressed and the bubble 
shape deviates from the ideal spherical one. 
We modify the code to track individual particles over the
full simulation. To separate trapped electrons, we introduce an energy
threshold: we assume that an electron was trapped if its maximum
energy exceeds $75~$MeV.

The distribution of the number of the electrons trapped in the bubble
as a function of their initial position $x_0$ is shown in
Fig.~\ref{fig2}(b) for three instants of time. Fig.~\ref{fig2}(b) demonstrates 
that the capture process is non-uniform and the number of trapped electrons changes 
along the laser path. The trapped electrons start motion from position $x=x_0$ at $t=t_0$ that  corresponds to the electron coordinates $y \approx R$,  $\xi = 0$ and 
 the bubble center position $x=x_0$ according to our model.
If the bubble propagates with velocity $V(t)$ then $x_0 = \int_0^{t_0} V(t')  dt'$. 
For the particle position 
we have $x = x_0 + \int_{t_0}^t v_x(t_0,t') dt'$, where $v_x$ is longitudinal 
component of the electron velocity and $v_x(t_0,t_0)=0$.
Differentiating this equation with respect to $t_{0}$ we get
$dx/dx_0 =  1 + [1/V(t_0)]\int_{t_0}^t [dv_x(t_0,t')/dt_0] dt'$, 
where we use $dx_0/dt_0 = V(t_0)$. In our model 
$v_x(t_0,t)$ is a function of $t-t_0$ and for a highly relativistic electron
$\gamma \gg \gamma _0 \gg 1$, $v_x \simeq 1$ we can write a
$dx/dx_0 \simeq - 1/(2\gamma_0^2)$. This derivation does not take into account the 
bubble shape deformation. However, it follows from our simulations that $\gamma _0$ calculated from 
$dx/dx_0 \simeq - 1/(2\gamma_0^2)$ largely determines the gamma-factor of the bubble back 
because the trapped electrons with $1<\gamma<\gamma_0$ spent most
of time in the bubble back. It is seen from Fig.~\ref{fig2}(b)-(d) that the self-injection is
 strongly correlated with this $\gamma-$factor. The number of  trapped particles 
per laser period peaks in the regions $1100 \lambda / c < t < 1400 \lambda / c$ and 
$t > 1700 \lambda / c $,  where $\gamma _0$ is minimal.  Vice versa, for 
$1400 \lambda / c < t < 1700 \lambda / c$ when $\gamma _0 >  2^{-1/2} R$ 
the number of the trapped particles becomes negligible that agrees 
with predictions of the model.

In Conclusion we present the model  for electron self-injection in the
relativistic plasma wakefield generated by a short laser pulse or by
an electron beam. The density threshold for electron self-injection 
is predicted by our model that is different from the self-injection condition proposed
in Ref.~\cite{Lu2007}. The evidences of such threshold have been 
observed in experiments \cite{Faure2006}. 
The model can also explain electron self-injection scheme based on 
downward plasma density transition \cite{Suk2001}. 
Such transition leads to the bubble elongation because of plasma density decreasing 
Further studies are neededand sharply reduces the gamma-factor of 
the bubble back thereby strongly enhancing self-injection. Recently, a strong impact 
of the wake phase velocity on electron trapping was also observed  
experimentally \cite{Fang2009} that is  consistent with our model. 
However, the accurate measurements of the bubble velocity and the 
structure of the electromagnetic field in the bubble are needed for
careful verification of the obtained results. 
 Much effort is now mounted in theory and in experiments 
to control the emittance and the energy spread of the accelerated electron bunch. It follows
from the obtained result that the self-injection strongly depends
on the wake phase velocity that can be controlled, for example, by
plasma density profiling \cite{Pukhov2008}. Therefore, the obtained
results can be used to optimize the plasma-based acceleration.

\begin{acknowledgments}
This work has been supported in parts by Russian Foundation for Basic Research
(Grant No 07-02-01239, 07-02-01265, 08-02-01209) and by DFG Transregio TR-18, Germany. 
\end{acknowledgments}

\end{document}